\documentclass[prl,twocolumn,amsmath,amssymb]{revtex4}

\usepackage{graphicx}

\begin{document}

\title{Generalized nonlocal optical response in nanoplasmonics}

\author{N.A.~Mortensen,$^{1,2}$ S.~Raza,$^{1,3}$ M.~Wubs,$^{1,2}$ T.~S{\o}ndergaard,$^{4}$ \& S.~I.~Bozhevolnyi$^5$}

\affiliation{
$^1$Department of Photonics Engineering, Technical University of Denmark, DK-2800 Kongens Lyng\-by, Denmark\\
$^2$Center for Nanostructured Graphene (CNG), Technical University of Denmark, DK-2800 Kong\-ens Lyngby, Denmark\\
$^3$Center for Electron Nanoscopy, Technical University of Denmark, DK-2800 Kongens Lyngby, Denmark\\
$^4$Department of Physics and Nanotechnology, Aalborg University, DK-9220 Aalborg, Denmark\\
$^5$Institute of Technology and Innovation, University of Southern Denmark, DK-5230 Odense, Denmark
}
\begin{abstract}
Metallic nanostructures exhibit a multitude of optical resonances associated with localized surface plasmon excitations. Recent observations of plasmonic phenomena at the sub-nano\-meter to atomic scale have stimulated the development of various sophisticated  theoretical approaches for their description. Here instead we present a comparatively simple semiclassical generalized nonlocal optical response (GNOR) theory that unifies quantum-pressure convection effects and induced-charge diffusion kinetics, with a concomitant complex-valued GNOR parameter. Our theory explains surprisingly well both the frequency shifts and size-dependent damping in individual metallic nanoparticles (MNPs) as well as the observed broadening of the cross-over regime from bonding-dipole plasmons to charge-transfer plasmons in MNP dimers, thus unraveling a classical broadening mechanism that even dominates the widely anticipated short-circuiting by quantum tunneling. We anticipate that the GNOR theory can be successfully applied in plasmonics to a wide class of conducting media, including doped semiconductors and low-dimensional materials such as graphene.
\end{abstract}

\maketitle

Studies of transport and wave dynamics in complex and confined geometries~\cite{vanRossum:1999} are now bridging several fields ranging from nanoplasmonics~\cite{Schuller:2010,Gramotnev:2010} and metamaterials~\cite{Engheta:2007} to molecular electronics~\cite{Nitzan:2003} and mesocopic quantum transport~\cite{Datta,Beenakker:1997}, with e.g. charge carriers responding to externally perturbing fields as well as exhibiting stochastical kinetics and entropic effects such as diffusion~\cite{Burada:2009}. When considering ultra-fast responses of optically driven collective plasma oscillations in nanoscale geometries, it is expected that the optical response should exhibit both quantum properties of the electron gas as well as classical diffusion dynamics of the optically induced charge. The co-existence and interplay of quantum and classical effects have profound implications for our understanding of light-matter interactions at the nanoscale, with direct relevance to the emerging field of quantum plasmonics~\cite{Tame:2013}.

\begin{figure}[b!]
\begin{center}
\includegraphics[width=0.89\columnwidth,clip]{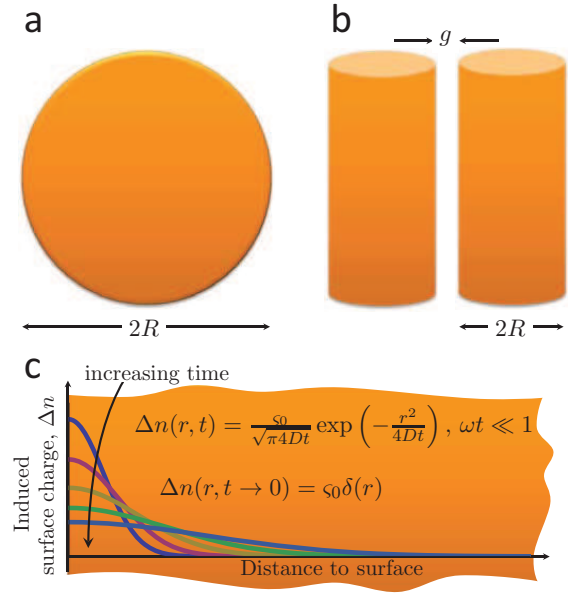}
\end{center}
\caption{Schematic illustration of nanoplasmonic geometries, and the observed properties that are explained by the generalized nonlocal optical response (GNOR) theory. {\bf a}, Spherical particle of radius $R$, known to exhibit size-dependent damping and resonant shifts. {\bf b}, Nanowires of radius $R$ arranged parallel to each other with a dimer gap $g$, known to exhibit gap-size-dependent broadening and shifts of hybridized plasmonic resonances. {\bf c}, Diffusive temporal spreading of an initially pure surface charge $\varsigma_0$ into a plasmonic nanoparticle. By accounting for diffusion of charge, the GNOR theory can explain size-dependent broadening and shifts of nanoparticle resonances, as well as gap-size-dependent broadening and shifts for dimer resonances. Unlike other theories, our GNOR theory does not invoke quantum tunneling to explain gap-size-dependent spectral broadening.\label{fig1}}
\end{figure}

The behavior of plasmon resonances of individual silver MNPs~\cite{Scholl:2012} and gold MNP dimers~\cite{Savage:2012} seems to be possible to understand only by invoking quantum-mechanical effects, i.e. quantum electron transitions and quantum tunneling, respectively. At the same time, one might question the necessity of considering numerous quantum-level transitions in nm-sized NPs (i.e., consisting of thousands of atoms and with a size much exceeding the Fermi wavelength) in the first case and the very possible existence of ultrafast tunneling phenomena (i.e., tunneling currents oscillating at optical frequencies) in the second case. While classical electrodynamics in a Drude local-response approximation (LRA) unambiguously fails explaining the observed phenomena, we show that the possibility for semiclassical accounts have not been exhausted.

Linear-response theory is inherent to our understanding of situations where matter is subject to externally perturbing fields. Common strategies assume a temporally instantaneous and spatially local response, while Nature is rich on examples where underlying degrees of freedom are responsible for a much more complex response. Materials exhibiting frequency dispersion are well-known for having complex-valued response functions due to Kramers--Kronig relations that originate from the ubiquitous principle of causality. By contrast, spatial dispersion can usually be neglected and most materials are well treated within LRA. Insulators represent a prime example since the polarization of one particular atom in the crystal is only weakly affected by coupling to neighboring atoms. Conducting media constitute a clear exception to this picture~\cite{Landau-Lifshitz-Pitaevskii,Boardman:1982a} and despite the widespread use of LRA approaches, the free carriers may mediate a response over finite distances that cannot necessarily be neglected in a nanoplasmonic context.

In terms of the Maxwell equations, the electrical field in a medium with nonlocal response is formally governed by
\begin{equation}\label{eq:nonlocal}
{\mathbf\nabla}\times{\mathbf\nabla}\times{\mathbf E}({\bf r})=\left(\tfrac{\omega}{c}\right)^2\int d{\bf r}'\, {\mathbf \varepsilon}({\bf r},{\bf r}') {\mathbf E}({\bf r}')
\end{equation}
where ${\mathbf \varepsilon}({\bf r},{\bf r}')$ is the nonlocal response function. This general concept of nonlocal response of conducting media originates from the competing mechanisms of pressure-driven convective flow of charge as well as disorder- or entropy-driven diffusion of charge~\cite{Landau-Lifshitz-Pitaevskii}. Quite surprisingly, while the literature is rich on discussions of the former effect within hydrodynamic models, the importance of the latter in nanoplasmonic systems remains unexplored, and, according to our knowledge, there is no unifying real-space description applicable to realistic plasmonic nanostructures. Pioneering works focused on pressure-driven convective flow of charge in ideal geometries~\cite{Fuchs:1969,Ruppin:1973,Dasgupta:1981,Garcia-de-Abajo:2008}, while the exploration of nonlocal response in arbitrarily shaped metal nanostructures has only recently been initiated~\cite{McMahon:2009}, emphasizing real-space rigorous formulations of semiclassical hydrodynamic equations~\cite{Raza:2011a} and different solution strategies~\cite{Toscano:2012a,Hiremath:2012,Toscano:2013,Luo:2013,Yan:2013}. Thus, large blueshifts in nanoscale noble metal plasmonic structures~\cite{Scholl:2012,Ciraci:2012,Raza:2013} have been interpreted in the context of the quantum-pressure related nonlocal response~\cite{Ciraci:2012,Raza:2013}, while quantum confinement~\cite{Scholl:2012} and surface-screening~\cite{Monreal:2013} explanations have also been proposed.

Here, we develop a semiclassical generalized nonlocal optical response (GNOR) theory that incorporates both quantum-pressure effects and induced-charge diffusion kinetics. We show that the GNOR approach can account for the main features observed in recent optical experiments with plasmonic nanostructures~\cite{Scholl:2012,Raza:2013,Savage:2012,Scholl:2013} without accounts for quantized-energy transitions and without invoking quantum tunneling that should not, as we argue later on, be important at optical frequencies.

We take equation~(\ref{eq:nonlocal}) as our starting point, while assuming a generic short-range isotropic response. Irrespective of the detailed microscopic mechanism behind the nonlocal response, the wave equation in the metal can then be reframed as~\cite{Mortensen:2013}
\begin{equation}\label{eq:wave-laplacian-phenomenological}
{\mathbf\nabla}\times{\mathbf\nabla}\times{\mathbf E}({\bf r})=\left(\tfrac{\omega}{c}\right)^2\left[\varepsilon_{\scriptscriptstyle D}+\xi^2\nabla^2\right]{\mathbf E}({\bf r}),
\end{equation}
where $\varepsilon_D$ is the Drude dielectric function usually associated with Ohmic local response of the electron gas (possibly generalized to also include interband effects), while the GNOR parameter $\xi$ represents a phenomenological length scale associated with the short-range nonlocal correction to the local-response Drude part. Importantly, the GNOR parameter accounts for mechanisms of very different origins that may compete or play in concert, while causing the same Laplacian-type correction to the LRA. For example, both convection and diffusion can lead to spatial dispersion in conducting media~\cite{Landau-Lifshitz-Pitaevskii}. Within LRA, the induced charge density $\Delta n$ is a delta function at the surface of the metal and diffusion will naturally smear this charge density with the short-time dynamics characteristic for pure diffusion (Fig.~\ref{fig1}c). Convection also tends to spread the charge density. In the following, we treat both dynamical effects on an equal footing by considering both propagating longitudinal pressure waves (in a hydrodynamic model) and diffusion (in convection-diffusion model). The main result of our analysis is the following expression for the GNOR parameter:
 \begin{equation}\label{eq:xi2}
\xi^2\simeq \frac{\beta^2}{\omega^2}-i\frac{D}{\omega}
\end{equation}
where $\beta\propto v_F$ is a characteristic velocity associated with pressure waves in the electron gas ($v_F$ being the Fermi velocity) while $D$ is the diffusion constant for the charge-carrier diffusion. The former is already known to cause frequency shifts (blueshifts)~\cite{Ruppin:1973,Raza:2011a,Toscano:2012a}, while the latter turns out to cause line-broadening, i.e. the GNOR parameter $\xi$ is in general a complex-valued quantity. As also anticipated from more general discussions~\cite{Landau-Lifshitz-Pitaevskii}, our rigorous semiclassical treatment shows that nonlocal effects may manifest themselves over distances greatly exceeding atomic dimensions and become comparable to characteristic structure dimensions, such as the radius $R$ of a nanoparticle (Fig.~\ref{fig1}a) or the gap distance $g$ in a dimer (Fig.~\ref{fig1}b). As a main result, we show that the GNOR even dominates more pure quantum-mechanical effects in the electromagnetic response at optical frequencies, such as the anticipated effect of quantum-mechanical tunneling currents in dimers with sub-nanometer gaps (Fig.~\ref{fig1}b).

\emph{Semiclassical response models.} We consider the standard equation-of-motion for an electron in an external electrical field subject to the continuity equation. The common LRA simply neglects effects of quantum pressure as well as the diffusion-contributions to the induced currents. We address these two aspects in turn and finally discuss how they play in concert to result in a complex-valued GNOR parameter.

Within the hydrodynamic model (including quantum pressure, but neglecting diffusion) the response is governed by a generalized constitutive equation~\cite{Raza:2011a,Toscano:2012a}
\begin{equation}\label{eq:eombeta}
  \frac{\beta^2}{\omega(\omega+i\gamma)}{\boldsymbol \nabla} \left({\boldsymbol \nabla}\cdot {\boldsymbol J}_{\scriptscriptstyle\rm conv}\right)+{\boldsymbol J}_{\scriptscriptstyle\rm conv} = \sigma_D{\boldsymbol E}
\end{equation}
where $\sigma_D$ is the usual Drude conductivity. Within Thomas--Fermi theory, $\beta^2 = (3/5)v_F^2$ and $\gamma=1/\tau$ is the damping rate also present in the Drude theory.

Taking the opposite standpoint (including diffusion, while neglecting quantum pressure), linearization of the problem gives 
\begin{equation}\label{eq:eomD}
  \frac{D}{i\omega}{\boldsymbol \nabla} \left({\boldsymbol \nabla}\cdot {\boldsymbol J}_{\scriptscriptstyle\rm dif}\right)+{\boldsymbol J}_{\scriptscriptstyle\rm dif} = \sigma_D{\boldsymbol E}
\end{equation}
as also derived recently in the context of metamaterial wire media~\cite{Hanson:2010}. Our key observation is that this result is mathematically similar to equation~(\ref{eq:eombeta}), while the different physical origins cause different prefactors for the nonlocal correction to Ohm's law.

Turning to the Maxwell wave equation the above nonlocal ${\boldsymbol \nabla} \left({\boldsymbol \nabla}\cdot {\boldsymbol J}\right)$ corrections to Ohm's law can be rewritten as a Laplacian correction to the Drude dielectric function~\cite{Toscano:2013,Mortensen:2013}, as anticipated in equation~(\ref{eq:wave-laplacian-phenomenological}). The convection and diffusion components of the current are of the same mathematical form and subject to the same boundary conditions. Thus, in the linear response the considered nonlocal contributions add up (as in semiconductor drift-diffusion theory), as confirmed by linearizing the full hydrodynamic diffusion-convection problem, revealing how both quantum and classical kinetic effects can play in concert contributing to the nonlocal response. Thereby we arrive at equation~(\ref{eq:wave-laplacian-phenomenological}) with
\begin{equation}\label{eq:xi2-b}
\xi^2=\frac{\beta^2}{\omega(\omega+i\gamma)}-i\frac{D}{\omega}=\frac{\beta^2+D(\gamma-i\omega)}{\omega(\omega+i\gamma)},
\end{equation}
which becomes Eq.~(\ref{eq:xi2}) when neglecting damping. Note that for higher $\omega$, diffusion becomes relatively more important compared to convection. Another important practical observation is that diffusion effectively causes the following modification of the nonlocal $beta$ parameter appearing in prior hydrodynamic work: $\beta^2\rightarrow \beta^2+D(\gamma-i\omega)$. We neglect electron spill-out and the associated boundary conditions remain unchanged in the presence of diffusion (${\boldsymbol n}\cdot{\boldsymbol J}=0$ on the metal surfaces so that no electrons escape the metal volumes). Thus existing numerical schemes and methods~\cite{Toscano:2012a,Hiremath:2012,Toscano:2013,Luo:2013,Yan:2013} can readily be exploited to implement the GNOR approach for various plasmonic configurations.

The diffusion constant is generally interlinked with other transport parameters such as the scattering time, i.e. $D\propto \tau=\gamma^{-1}$. For $\omega\gg\gamma$, we thus recover equation~(\ref{eq:xi2}) for the GNOR parameter that now explicitly exposes the two competing length scales previously discussed only qualitatively~\cite{Landau-Lifshitz-Pitaevskii}, i.e. the convection length $\beta/\omega$ on the one hand and the diffusion length $\sqrt{D/\omega}$ on the other hand. The damping associated with the latter is an important new finding which turns out to be crucial when approaching the nanoscale. Diffusion degrades plasmonic excitations, providing an additional broadening mechanism that, mathematically, is enacted by an imaginary contribution to $\xi^2$.

\emph{Validity domain.} For classical gases (such as dilute plasmas, electrolytes, and weakly doped semiconductors), the velocity distribution is governed by Maxwell--Boltzmann statistics and $D$ is proportional to the temperature, as given by the Einstein relation. For metals on the other hand, Fermi--Dirac statistics implies a narrow transport velocity distribution~\cite{Landau-Lifshitz-Pitaevskii} with a characteristic velocity $v_F$. As a result, the diffusion constant is simply $D\simeq v_F^2 \tau$, corresponding to a mean-free path of $\ell=v_F \tau$. We point out that our diffusive model is valid for structural dimensions exceeding the mean-free path that in pure single crystals~\cite{Sondheimer:1952} can be of the order of 100\,nm for Ag and Au, down to approximately 3\,nm for Na. Moreover, in realistic plasmonic nanostructures, $\ell$ depends on actual material processing conditions, becoming shorter than in single-crystalline bulk metals. This enlarges the validity domain of a diffusion description to include structures with dimensions of only a few nanometers. For even smaller dimensions, electrons will move ballistically between the surfaces of the structure, and surface scattering might become important. For metals, equation~(\ref{eq:xi2-b}) simplifies to $\xi^2= \frac{v_F^2}{\omega^2}\left(\tfrac{3}{5} -i\tfrac{\omega}{\gamma}\right)+{\mathcal O}(\gamma/\omega)$. This result leads to an important insight into the interplay of different broadening mechanisms: the lower the Ohmic loss and absorption, the more important is the nonlocal response due to long-range diffusion of the induced charge. Re-introducing the diffusion constant and the \emph{beta} parameter we arrive at equation~(\ref{eq:xi2}) which holds if the mean-free path significantly exceeds the convective length, i.e. $\ell\gg v_F/\omega$.

We provide in the following two key examples of the GNOR approach, demonstrating that the interplay of quantum pressure and diffusion has a remarkable impact on the optical response of plasmonics nanostructures and solving longstanding open problems.

\begin{figure*}[t!]
\begin{center}
\includegraphics[width=1.5\columnwidth]{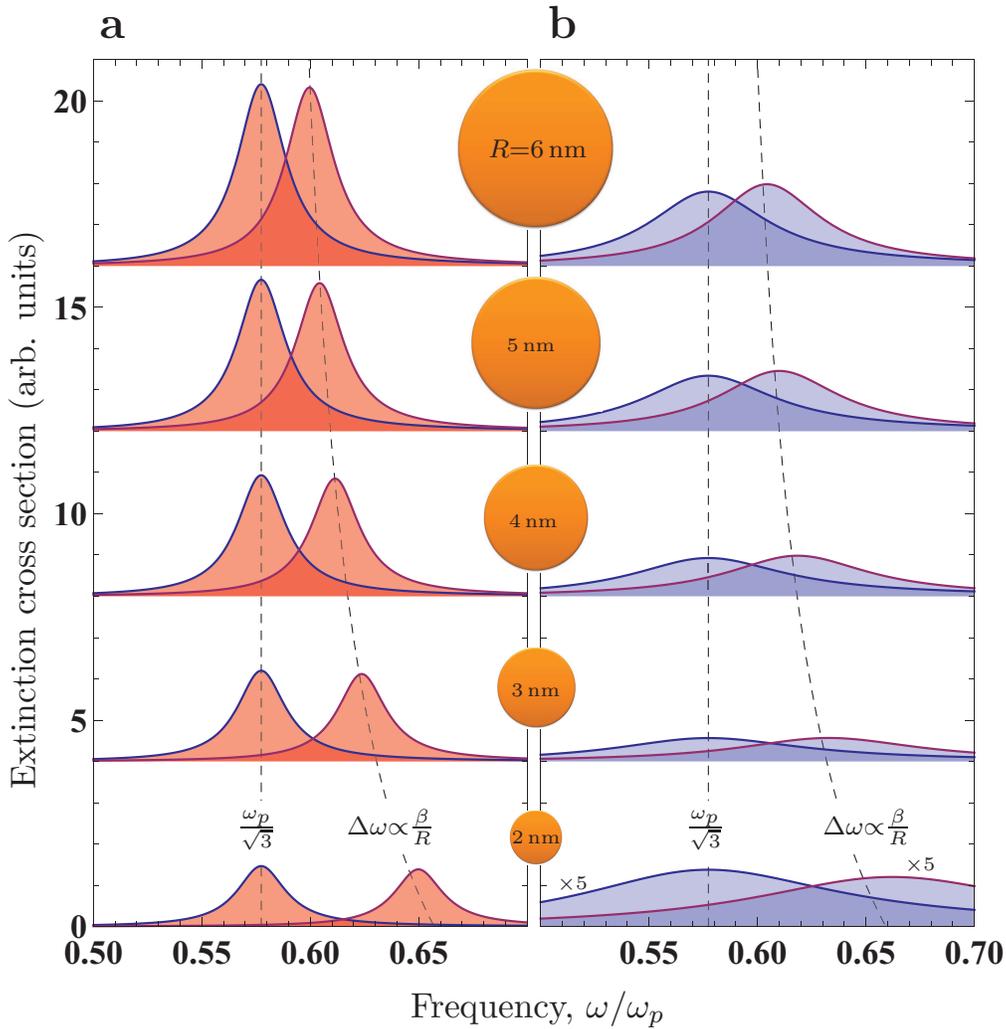}
\end{center}
\caption{Extinction cross section, blueshift, and size-dependent damping for a metal sphere. The sphere radius $R$ is varied from 2\,nm to 6\,nm. {\bf a}, Local-response approximation ($\beta=D=0$) versus hydrodynamic nonlocal response ($\beta\neq 0$, $D=0$). {\bf b}, Local-response approximation (including $1/R$ Kreibig damping) versus the GNOR model ($\beta\neq 0$, $D\neq 0$). The metal parameters for sodium are used~\cite{Teperik:2013a,Kreibig:1995}: $\omega_\text{p}=5.89$~eV, $\gamma=0.16$~eV, $v_\text{F}=1.05\times10^6$~m/s, $\beta=0.81\times10^6$~m/s, $D=2.04\times10^{-4}\,$m$^2$/s, and $A=1$.\label{fig2}}
\end{figure*}

\emph{Size-dependent damping.} In general, the hydrodynamic corrections give a blueshift of resonances as the characteristic dimensions are reduced~\cite{Ruppin:1973,Toscano:2012a,Ciraci:2012,Raza:2013}. With the complex-valued GNOR parameter $\xi$ at hand, we now anticipate the blueshift to occur \emph{along} with broadening of the resonant response when decreasing characteristic structure dimensions. In the case of a spherical particle, the blueshift has a $\beta/R$ dependence~\cite{Raza:2013}, leading us to foresee that the line broadening scales as $1/R$ as well. In the quasi-static limit ($\lambda\gg R$), one can straightforwardly work out the complex-valued resonance frequency $\omega=\omega'+i\omega''$ by considering the polarizability pole. As usual, the real part $\omega'$ gives the surface plasmon (SP) resonance frequency, while the imaginary part $\omega''$ is related to the resonance linewidth. For simplicity, we consider the case of a particle in vacuum with no interband effects and find (to second order in $1/R$)
\begin{subequations}
 \label{eq:resfreq}
\begin{align}
    \omega' &\simeq  \frac{ \omega_\text{p} }{\sqrt{3}} + \frac{\sqrt{2}\beta}{2R}, \\
    \omega'' &\simeq  -\frac{\gamma}{2}- \frac{\sqrt{6}}{24} \frac{D\omega_\text{p}}{\beta R}.
\end{align}
\end{subequations}
It transpires clearly that the $1/R$ size-dependent nonlocal effects are present in both the resonance frequency and linewidth. It should be emphasized that, until now, line shifts have been explained by nonlocal response (and competing theories), whereas the line broadening was 'put in by hand'. Here, by using the GNOR theory we have arrived at a unified explanation of both experimentally observed phenomena by nonlocal effects. Line broadening has been seen experimentally in the extinction of small particles~\cite{Kreibig:1969,Kreibig:1995,Gaudry:2003,Scaffardi:2006,Kolwas:2013} and EELS measurements on plasmons in thin nanowires and bow-tie antennas have also revealed plasmon losses exceeding the expectations based on bulk-damping parameters~\cite{Nicoletti:2011,Wiener:2013}. In the literature such line broadening has often been phenomenologically accounted for by a size-dependent damping rate~\cite{Kreibig:1969,Kreibig:1995,Gaudry:2003,Scaffardi:2006,Kolwas:2013}, but without placing it in the context of nonlocal semiclassical equations of motion.

The phenomenology introduced by Kreibig~\cite{Kreibig:1969,Kreibig:1995} describes the linewidth broadening by introducing a size-dependent correction to the damping rate: $\gamma\rightarrow \gamma + A v_F/R$. Equipped with the GNOR theory, one does not need to assume the $1/R$-dependence: it comes immediately out as a consequence of the GNOR correction to the dipolar sphere polarizability. By comparison to the Kreibig model we formally find that $A= \sqrt{\frac{1}{24}} \frac{D\omega_\text{p}}{\beta v_F}$. For metals this implies that $A\sim \omega_p\tau/4$. Use of bulk values for noble metals would estimate a too high $A$ parameter compared to experiments where for spheres $A$ is estimated to be of the order unity~\cite{Kreibig:1969,Kreibig:1995}.

So far, we have assumed that the transport time is given by the bulk relaxation time $\tau_0$. However, in the experiments dealing with plasmonic nanostructures, it has been found that one has to increase the collision frequency~\cite{Zhang:2005,Liu:2010} or the imaginary part of permittivity~\cite{Pors:2013} by several times as compared to the bulk metal (gold) parameters in order for the simulations to better correspond to the experimental observations. This \emph{size-independent} correction factor was ascribed to the influence of the  surface scattering and grain boundary effects in nanostructures~\cite{Zhang:2005,Liu:2010,Pors:2013}. In our case, we can simply introduce the characteristic relaxation time $\tau_s$ associated with these effects, which can be estimated from the condition that $A\sim 1$ as $\tau_s\sim 4/\omega_p$.

With the above refinement at hand we can demonstrate using numerical simulations (Fig.~\ref{fig2}) that, in agreement with equations~(\ref{eq:resfreq}), the hydrodynamic response causes a blueshift (Fig.~\ref{fig2}a) whereas the diffusion causes an additional broadening (Fig.~\ref{fig2}b). In the latter figure, it is also shown that a similar resonance broadening (but not the blueshift) is predicted by the LRA with additional Kreibig damping.

Our real-space nonlocal wave equation (\ref{eq:wave-laplacian-phenomenological}) along with the GNOR parameter (\ref{eq:xi2}), which unravels the fundamental link between diffusive broadening and the Kreibig-like surface scattering, enables thereby one to solve a longstanding open problem: surface-related scattering can now be computationally accounted for also in complex-shaped geometries beyond that of spherical symmetry and low radius of curvature.

\emph{Nanowire dimers.} Plasmonic dimers (Fig.~\ref{fig1}b) are rich on hybridization phenomena as the gap distance $g$ is reduced~\cite{Prodan:2003} and nonlocal hydrodynamic effects on both hybridization and field enhancement have been anticipated~\cite{Toscano:2012a}. In order to elucidate the diffusion contribution to nonlocal effects, we may consider dimers of nanowires where the nanowire radius $R$ itself is too large to cause either nonlocal hydrodynamic effects or increased damping of the Kreibig kind. Nevertheless, we expect that the dimer would exhibit nonlocal effects once the dimer gap distance $g$ turns comparable in magnitude to $\xi$, resulting in additional broadening in the vicinity of the gap as $v_F/g$ increases. With our present formalism this can now be quantified without any need to invoke {\em ad hoc} assumptions specifically for dimers.

\begin{figure}[t!]
\begin{center}
\includegraphics[width=1\columnwidth]{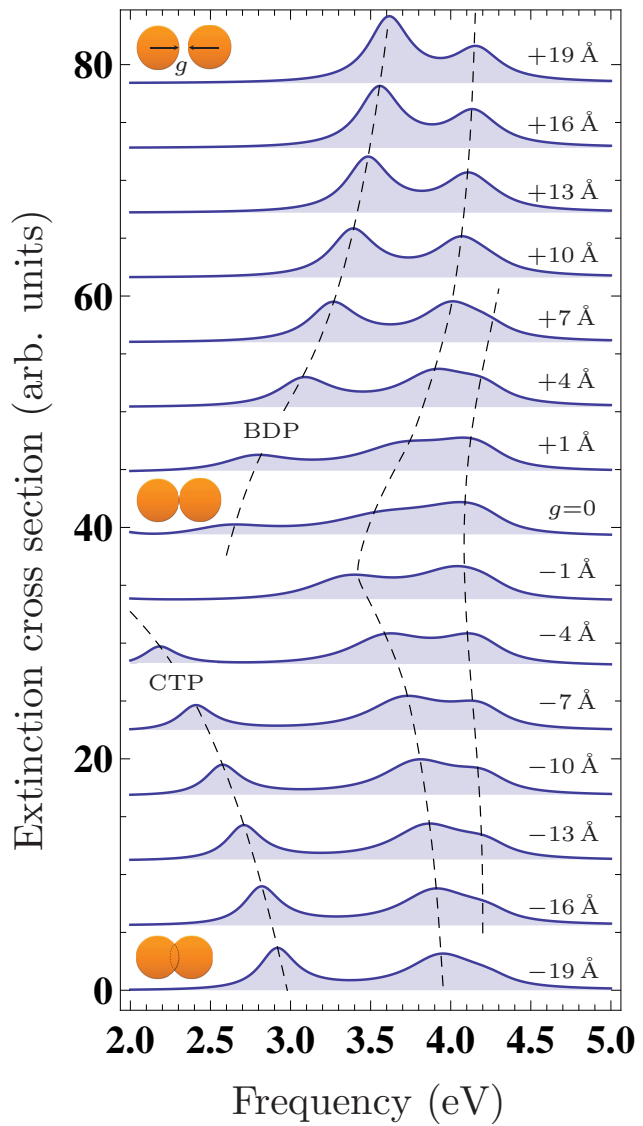}
\end{center}
\caption{Extinction cross section for a nanowire dimer with a sub-nanometer gap within the GNOR model. The radius of the sodium wires is $R=4.9$\,nm with the gap $g$ varying from $-19\,${\AA} to $+19\,${\AA}. The diffusion constant $D=1.36\times10^{-4}\,$m$^2$/s causes GNOR spectra in accordance with the TD-DFT calculations by Teperik~\emph{et al.} (Ref.~\onlinecite{Teperik:2013a}) and in overall agreement with the broadening observed experimentally by Savage \emph{et al.} (Ref.~\onlinecite{Savage:2012}).\label{fig3}}
\end{figure}

Although diffusion is of a classical origin, the discussion of its effect in dimers ties up with very recent experiments on dimers in the quantum-tunneling regime~\cite{Savage:2012}. \emph{Ab-initio} approaches show a crossing from the classical hybridization of localized surface-plasmon resonances to tunneling-mediated charge-transfer plasmons~\cite{Stella:2013,Teperik:2013a,Andersen:2013,Teperik:2013b}. Being able to push experiments into this intriguing re\-gime~\cite{Savage:2012,Scholl:2013,Kadkhodazadeh:2013}, commonly associated with expectations of quantum physics, leaves an open question: \emph{Can this regime be adequately described with semiclassical models?} While nonlocal response within the hydrodynamic semiclassical model has been found unsuccessful in explaining features from time-dependent density-functional theory (TD-DFT)~\cite{Stella:2013,Teperik:2013a,Teperik:2013b} there have been phenomenological attempts of classically modeling the cross-over regime. The 'quantum-corrected model' (QCM)~\cite{Esteban:2012} adds an artificial conducting and lossy material in the gap to mimic short-circuiting currents associated with quantum tunneling. While apparently successful in qualitatively fitting results of \emph{ab-initio} simulations~\cite{Esteban:2012,Teperik:2013a,Teperik:2013b}, the model raises concerns regarding its physical foundation. The well-established understanding of mesoscopic quantum-electron transport~\cite{Datta} is that the tunneling through the classically forbidden gap region is elastic (ballistic transport) while energy relaxation takes place inside the metallic contact regions. Opposite to that, the artificial gap material introduced in the QCM causes dissipation \emph{within} the gap, while there is no associated relaxation occurring on the metal-sides of the junction.

While \emph{ab-initio} works emphasize tunneling~\cite{Esteban:2012,Teperik:2013a,Teperik:2013b}, recent experiments on dimers~\cite{Savage:2012,Scholl:2013} do not offer explicit evidence that the broadening is associated with quantum tunneling. The formation of a sub-nanometer gap is evident from the observed DC voltage-driven tunneling current~\cite{Savage:2012}, while there is no explicit confirmation of AC tunneling currents caused by the optical driving. Earlier time-resolved tunneling experiments have reported tunneling RC times in the picosecond range~\cite{Weiss:1995}, thus suggesting a suppression of optical AC tunneling currents that would take place in femtoseconds. This apparently makes quantum tunneling dynamics too slow and a less likely mechanism to explain the broadening of dimer modes at optical frequencies. Applying the GNOR framework, we demonstrate in the following that the diffusion offers a strong competing damping mechanism. In fact, for fast driving of the junction~\cite{Grifoni:1998}, diffusion may completely dominate the dissipation of the dimer junction as we illustrate by a circuit analysis (see Supplementary Information).

The diffusion-driven damping occurs right inside the surface of the metals (not in the gap), becoming progressively more pronounced for smaller gaps and vanishing for large gaps. To exemplify this in detail we revisit recent nonlocal (hydrodynamic) simulations~\cite{Teperik:2013a}, while making sure to account for the diffusive broadening as well by use of the complex-valued GNOR parameter (Fig.~\ref{fig3}). For relatively large gaps, one observes the bonding-dipole plasmon (BDP) along with higher-order modes known to appear below the plasma frequency within the LRA~\cite{Romero:2006} (for a larger radius this becomes particularly clear, not shown). As the gap is reduced to the nonlocal regime $g\lesssim v_F/ \omega$, as considered in Fig.~\ref{fig3} for $R=4.9$\,nm, resonances are slightly blueshifted with respect to the LRA result~\cite{Toscano:2012a}. When the gap shrinks further, progressively stronger hybridization~\cite{Prodan:2003} and accordingly larger BDP redshifts are clearly seen. At the same time, the BDP is gradually suppressed due to the increasing role of diffusion as the contact point, $g=0$, is approached. This is in strong contrast to predictions from both the LRA~\cite{Romero:2006} and from previous nonlocal theories that treated the \emph{beta} parameter real-valued~\cite{Toscano:2012a,Fernandez-Dominguez:2012,Teperik:2013a,Teperik:2013b}. As we enter the contact regime, the BDP fades away, vanishing completely for $g<0$. For touching wires, the charge-transfer plasmon (CTP) appears, whose resonance blueshifts and grows in strength for larger wire overlaps. We note that, for $g\sim 0$, the diffusive broadening is so strong that only higher-order modes persist (as the induced surface charge is located away from the contact point), while both the BDP and the CTP are strongly suppressed. This makes a discussion on their possible co-existence problematic~\cite{Kadkhodazadeh:2013}. Finally, we note that in the anticipated tunneling regime the extinction spectra are strongly broadened by the complex nonlocal response. In fact, our semiclassical approach is in remarkable agreement with the TD-DFT results~\cite{Stella:2013,Teperik:2013a,Teperik:2013b}, with the diffusion contribution being responsible for 'repairing' the apparent incompatibility of TD-DFT calculations and earlier hydrodynamic predictions~\cite{Teperik:2013a,Teperik:2013b}. Our semiclassical GNOR theory thereby pinpoints induced-charge diffusion as the dominant broadening mechanism in recent experiments on plasmonic dimers~\cite{Savage:2012}, thus challenging tunneling-current interpretations for which the phenomenological QCM was constructed.

In this article we have presented a semiclassical (GNOR) approach that is offering a long-sought unification of nonlocal response mechanisms having both quantum-mechanical and classical origins. The GNOR theory places established observations of size-dependent damping into the context of nonlocal response and offers an accurate classical explanation of spectral broadening in metallic nanoparticle dimers without invoking quantum-mechanical tunneling, whose efficiency at optical frequencies is questionable. We have so far considered degenerate electron systems such as metals, where screening is strong and nonlocal effects manifest themselves in the nanometer to sub-nanometer regime. In the search for a new mesoscopic regime, where plasmons potentially exhibit both semiclassical dynamics and quantum effects, low-density doped semiconductors and tunable low-dimensional materials (including the graphene family of 2D materials) appear attractive~\cite{Britnell:2013}. Quantum light-matter interactions~\cite{Koppens:2011} and nonlocal response~\cite{Wang:2013} were already considered for graphene plasmonics, and, for such non-degenerate systems, our theory anticipates \emph{temperature-dependent} nonlocal response that might lead to novel nonlocal effects accessible via experimental observations.

\emph{Acknowledgments.} We acknowledge Giuseppe Toscano for use of his finite-element code. The Center for Nanostructured Graphene (CNG) is funded by the Danish National Research Foundation, Project DNRF58. The A. P. M{\o}ller and Chastine Mc-Kinney M{\o}ller Foundation is gratefully acknowledged for the contribution toward the establishment of the Center for Electron Nanoscopy. N.A.M. and M.W. acknowledge financial support by Danish Council for Independent Research - Natural Sciences, Project 1323-00087. We thank J{\o}rn Hvam and Antti-Pekka Jauho for stimulating discussions.

\end{document}